\documentclass[iop,apjl,numberedappendix,appendixfloats,twocolappendix,revtex4]{emulateapj} 
\usepackage{times} 
\usepackage{latexsym,amsmath,amssymb,amsfonts} 
\usepackage{graphicx} 
\usepackage{fancyhdr}  
\usepackage{natbib} 
\usepackage{rotfloat} 
\usepackage{verbatim} 
\usepackage{rotating}   
\usepackage{url}
\usepackage{dcolumn}
\usepackage{morefloats} 
\usepackage[breaklinks,colorlinks,citecolor=blue,linkcolor=magenta]{hyperref}  
\newcommand{\sqcm}{cm$^{-2}$}  
\newcommand{\lya}{Ly$\alpha$}
\newcommand{\lyb}{Ly$\beta$}
\newcommand{\lyg}{Ly$\gamma$}

\newcommand{\HI}{\mbox{H\,{\sc i}}}

\newcommand{\OVI}{\mbox{O\,{\sc vi}}}

\newcommand{\CII}{\mbox{C\,{\sc ii}}}

\newcommand{\CIII}{\mbox{C\,{\sc iii}}}

\newcommand{\NIII}{\mbox{N\,{\sc iii}}}

\newcommand{\NeVIII}{\mbox{Ne\,{\sc viii}}}

\newcommand{\zabs}{$z_{\rm abs}$}

\newcommand{\Msun}{$M_{\odot}$} 

\newcommand{\be}{\begin{equation}}
\newcommand{\en}{\end{equation}}

\def\zabs{$z_{\rm abs}$}

\def\kms{km~s$^{-1}$}

\shorttitle{Cool gas in cluster outskirts} 
\shortauthors{Muzahid et al.}
\begin{document}

\title{Discovery of an \HI-rich gas reservoir in the outskirts of SZ-effect selected clusters} 

\author{Sowgat Muzahid\altaffilmark{1},  
Jane Charlton\altaffilmark{2}, 
Daisuke Nagai\altaffilmark{3},  
Joop Schaye\altaffilmark{1}, and   
Raghunathan Srianand\altaffilmark{4}  
}    

\affil{\\ $^{1}$Leiden Observatory, Leiden University, PO Box 9513, NL-2300 RA Leiden, The Netherlands (\color{blue} sowgat@strw.leidenuniv.nl\color{black}) \\ 
$^{2}$Department of Astronomy \& Astrophysics, The Pennsylvania State University, State College, PA 16801, USA \\ 
$^{3}$Department of Physics, Yale University, New Haven, CT 06520, USA \\ 
$^{4}$Inter-University Centre for Astronomy and Astrophysics, Post Bag 4, Ganeshkhind, Pune 411 007, India \\ 
}  


\begin{abstract}

\noindent 
We report on the detection of three strong \HI\ absorbers originating in the outskirts (i.e., impact parameter, $\rho_{\rm cl}\approx$~(1.6--4.7)$r_{500}$) of three massive ($M_{500}\sim3\times10^{14}$~\Msun) clusters of galaxies at redshift $z_{\rm cl}\approx$~0.46, in the {\it Hubble Space Telescope} Cosmic Origins Spectrograph ($HST$/COS) spectra of 3 background UV-bright quasars. These clusters were discovered by the 2500 deg$^2$ South Pole Telescope Sunyaev--Zel'dovich (SZ) effect survey. All three COS spectra show partial Lyman limit absorber with $N(\HI)>10^{16.5}$~\sqcm\ near the photometric redshifts ($|\Delta z/(1+z)| \approx 0.03$) of the clusters. The compound probability of random occurrence of all three absorbers is $<0.02$\%, indicating that the absorbers are most likely related to the targeted clusters. We find that the outskirts of these SZ-selected clusters are remarkably rich in cool gas compared to existing observations of other clusters in the literature. The effective Doppler parameters of the Lyman series lines, obtained using single cloud curve-of-growth (COG) analysis, suggest a non-thermal/turbulent velocity of a few$\times10$~\kms\ in the absorbing gas. We emphasize the need for uniform galaxy surveys around these fields and for more UV observations of QSO-cluster pairs in general in order to improve the statistics and gain further insights into the unexplored territory of the largest collapsed cosmic structures.  

\end{abstract}


\section{Introduction} 
\label{sec:intro}

Galaxy clusters are the most massive gravitationally bound structures in the universe. With 100--1000 galaxies and total masses of $\sim$$10^{14-15}$~\Msun, gas accreting onto a cluster is typically heated to a very high temperature. In fact, X-ray observations have revealed enormous quantities of diffuse, hot ($\sim$$10^{7-8}$~K) gas in the central regions of galaxy clusters within which the mean mass density is over 500 times the critical density of the universe (i.e., $<r_{500}$; see \cite{Voit05}, for a review). The origin of the energy radiated away via X-rays, which have been the main source of information on the ICM so far, is thermal bremsstrahlung with power, $P\propto n_e^2 T_e^{1/2}$. However, the ICM in cluster outskirts (i.e., $\rho_{\rm cl}\sim$~(1--5)$r_{500}$), where the density and temperature are considerably lower than in the core, is not bright enough to detect in X-ray emission. Consequently, the outskirts of galaxy clusters, particularly at high-$z$, are not well explored observationally. This is partly because of the lack of sensitive diagnostics for probing the cool/warm-hot gas, with $T$$\sim$$10^{4-6}$~K, that prevails in the circumcluster medium (CCM).

In recent years, with the advent of high resolution cosmological simulations and deep X-ray observations of a handful of nearby clusters, cluster outskirts have emerged as one of the new frontiers of study in cluster astrophysics and cosmology \citep[e.g.,][]{Simionescu11,Walker12,Urban14,Lau15,Bahe17}. This environment is the interface between clusters and the cosmic web. In the outskirts, galaxies and groups of galaxies are stripped of their metal-rich gas by tidal forces and by the ram pressure provided by the cluster, enriching the ICM with heavy elements. The outskirts of galaxy clusters may harbor a substantial fraction of the ``missing baryons" \citep[e.g.,][]{Gonzalez07,Gonzalez13} which could reside in the cool/warm-hot ($T\sim$$10^{4-6}$~K) gas phase. Probing the CCM is thus crucial for understanding gas flows, metal enrichment history, and the baryon budget in the largest collapsed environments.       

Since cluster outskirts are beyond the reach of present day X-ray telescopes, an effective alternative is to use absorption line spectroscopy of background UV-bright quasars to probe the CCM. This technique has provided a wealth of information regarding the circumgalactic medium (CGM) of both low- and high-$z$ galaxies \citep[e.g.,][]{Turner14,Werk14,Kacprzak15}. However, except for a very few studies \citep[i.e.,][]{Yoon12,Yoon17,Burchett17}, it has not yet been used to probe the CCM. \cite{Yoon12} have studied 43 \lya\ absorbers along 23 background QSO-sightlines towards the Virgo cluster using COS, STIS, and GHRS data. Interestingly, they found that the cool gas in Virgo is preferentially located in the cluster outskirts and is associated with substructures. Recently, \cite{Burchett17} have studied the CCM of 7 X-ray detected clusters with masses of $M_{200} \sim$~few$\times10^{14} M_{\odot}$. Their sightlines typically pass within 300~kpc of a cluster galaxy. They have reported a very low covering fraction ($\approx$~18\%) of \HI\ absorbing gas (equivalent width $>30$~m\AA) in the CGM of cluster galaxies as compared to field/group galaxies ($\approx$~80--100\%).

\begin{table*}	
\begin{center}  
\caption{Details of the QSO-cluster pairs}   
\begin{tabular}{crccccccccccc}    
\hline \\ [-1.5ex]  
Cluster    &  RA$_{\rm cl}$  &  DEC$_{\rm cl}$     & $z_{\rm cl}$ &   $M_{500}$  & $r_{500}$ &  QSO  & $z_{\rm qso}$  &  $FUV$ &   $\rho_{\rm cl}$   &  
$\rho_{\rm cl}/r_{500}$  & \zabs  &  $\log N(\HI)$ \\         
           &  (J2000) &  (J2000) &              & $(10^{14}M_{\odot})$ &  (Mpc)&       &                           &         &  (Mpc)  &     &  &  ($N/\rm cm^{-2}$)  \\  
  (1)      &   (2)    &   (3)    &   (4)   &   (5)  &  (6)  &  (7)  &   (8)  &  (9)  &  (10)  &   (11)  &  (12)  & (13) \\   
\hline \\ [-1.5ex] 
J0041--5107 &  10.2932 & $-$51.1286 &  $0.45\pm0.04$  &    3.04$\pm$0.87 & 0.87 &  J0040--5057&  0.608&  17.43&  3.80  &    4.4  & 0.43737  & $18.63\pm0.07$ \\  
J2016--4517 & 304.0050 & $-$45.2978 &  $0.45\pm0.03$  &    3.19$\pm$0.89 & 0.89 &  J2017--4516&  0.692&  17.81&  4.20  &    4.7  & 0.43968  & $16.52\pm0.05$ \\   
J2109--5040 & 317.3825 & $-$50.6765 &  $0.47\pm0.04$  &    3.81$\pm$0.87 & 0.93 &  J2109--5042&  1.262&  17.93&  1.47  &    1.6  & 0.51484  & $16.68\pm0.03$ \\ \\ [-1.5ex]   
\hline   
\end{tabular}                                                         
\end{center} 
\vskip-0.2cm     
{\em Notes--} Cluster's name (column 1), right ascension (column 2), declination (column 3), photometric redshift (column 4), and mass (column 5) are from \citet{Bleem15}. The $r_{500}$ values are listed in column 6. QSO's name (column 7), emission redshift (column 8), and FUV magnitude (column 9) are from \cite{Monroe16}. The impact parameters and normalized impact parameters of the QSO sightlines are listed in columns 10 \& 11, respectively. The absorption redshifts and the \HI\ column densities measured from COS data are listed in columns 12 \&  13, respectively. The $N(\HI)$ values are obtained using single cloud COG analysis of Lyman series lines. These values are consistent with the ones obtained from the Lyman limit breaks. We note that the errors in the column densities are underestimated since we do not take the continuum placement uncertainties into account. A more realistic error would be 0.10--0.15 dex.               
\label{tab:tab1}  
\end{table*}   	

Motivated by the lack of UV observations of the CCM and its importance, we have built a sample of QSO-cluster pairs by cross-correlating the SZ-effect selected cluster catalog of \citet{Bleem15} and the all-sky UV-bright QSO catalog (UVQS) of \citet{Monroe16}. As a pilot program we have obtained far-UV (FUV) spectra of 3 quasars using $HST/$COS. These quasars probe the outskirts of 3 SZ-selected clusters of masses $M_{500}\sim3\times10^{14}$~\Msun\ at redshift $z_{\rm cl}\approx0.46$ with impact parameters, $\rho_{\rm cl}$, of 1.5--4.2~Mpc ($\rho_{\rm cl}/r_{500}\approx$~1.6--4.7). The details of the QSO-cluster pairs are listed in Table~\ref{tab:tab1}. Intriguingly, in all three cases we detect strong \HI\ absorption with $N(\HI)>10^{16.5}$~\sqcm\ at the redshifts of the foreground clusters.


This {\em Letter} is organized as follows: In Section~\ref{sec:observations} we describe our COS observations. The analysis and the main results are presented in Section~\ref{sec:analysis}. The possible implications of our observations are discussed in Section~\ref{sec:discuss}. Throughout the {\em Letter} we adopt a flat $\Lambda$CDM cosmology with $H_0=71$~\kms~$\rm Mpc^{-1}$, $\Omega_{\rm M} = 0.3$, and $\Omega_{\Lambda}=0.7$. All the distances given are proper (physical) distances.

\section{Observations}      
\label{sec:observations}   

UV spectra of the three background UV-bright ($FUV<18$) quasars were obtained using $HST/$COS Cycle-24 observations under program ID: GO-14655 (PI: Muzahid). The properties of COS and its in-flight operations are discussed in \citet{Osterman11} and \cite{Green12}. The observations consist of G130M and G160M FUV grating integrations covering 1100--1800~\AA\ at a medium resolution of $R\sim$18,000\footnote{No flux is observed below 1315~\AA\ in the spectrum of QSO UVQS~J0040--5057 due to the strong Lyman limit break in the spectrum caused by the \zabs~$=$~0.43737 absorber studied here.}. The data were retrieved from the $HST$ archive and reduced using the STScI CalCOS v3.1.8 pipeline software. The reduced, flux calibrated individual exposures were aligned and coadded using the {\sc idl} code ``coadd\_x1d''(v3.1) developed by \cite{Danforth10}. The combined spectra have a signal-to-noise ratio (S/N) of 5--10 per resolution element. Each combined spectrum was binned by 3 pixels since the COS FUV spectra, with six raw pixels per resolution element, are highly oversampled. The analysis/results presented here are not affected by this re-binning. Continuum normalizations were done by fitting the line-free regions with smooth low-order polynomials.

\begin{figure*} 
\centerline{\vbox{
\centerline{\hbox{ 
\includegraphics[width=0.75\textwidth,angle=270]{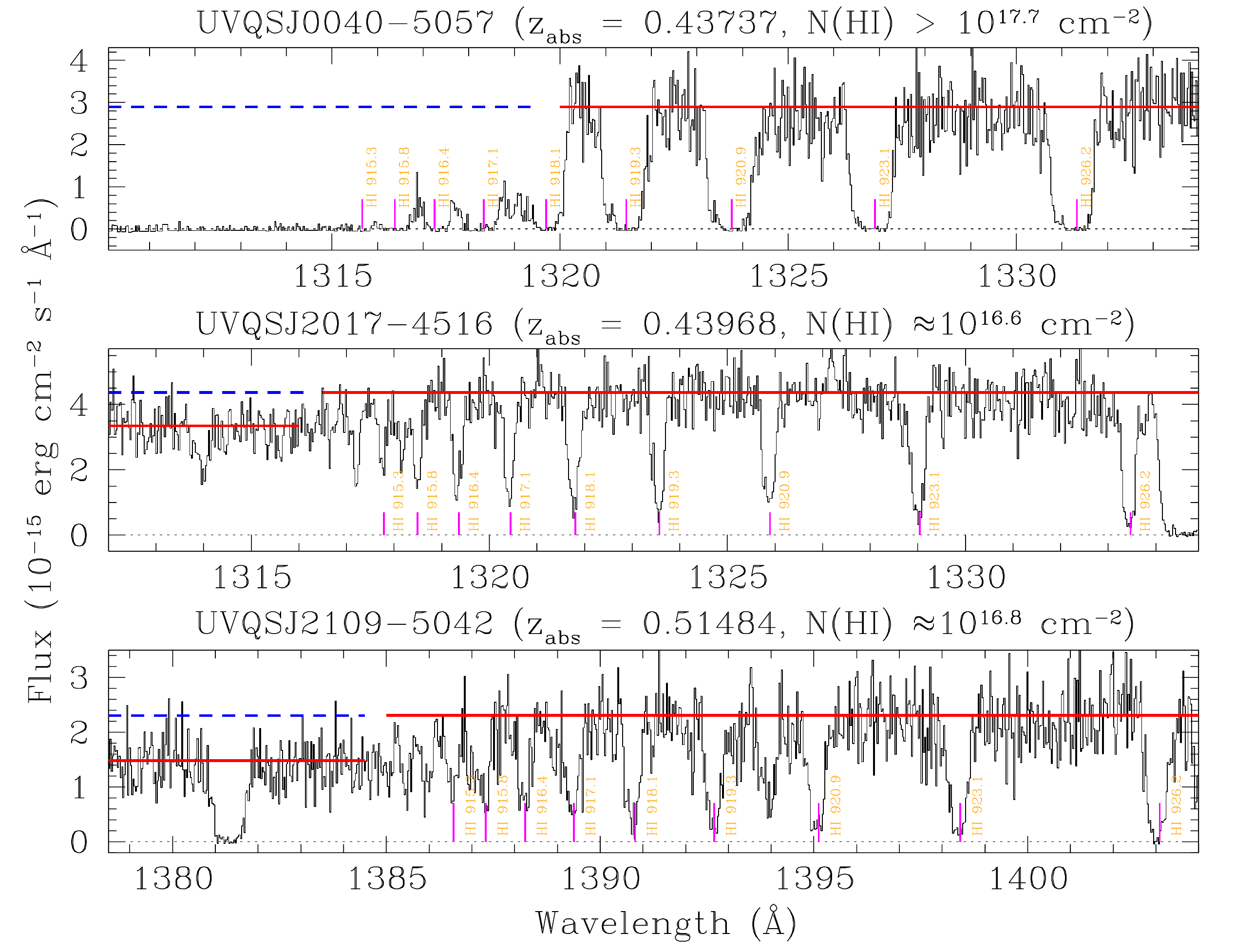}  
}} 
}}  
\vskip-0.2cm  
\caption{Selected parts of the COS spectra showing the Lyman limit breaks caused by the absorbers with redshifts consistent with the photometric redshifts of the targeted clusters. The higher order \HI\ Lyman series lines are marked. The horizontal (red) lines are the adopted continua. The horizontal (blue) dashed lines bluewards the Lyman limit breaks are the extrapolated continua.  The full break produced by the absorber at \zabs~$=0.43737$ towards UVQS~J0040--5057 only provides a lower limit on $N(\HI)$ of $10^{17.7}$~\sqcm. The partial breaks seen towards UVQS~J2017--4516 and UVQS~J2019--5042 yield $\log N(\HI)/\rm cm^{-2} \approx$~16.6 and 16.8, respectively. These $N(\HI)$ values are consistent with the ones obtained using single cloud COG analysis.}         
\label{fig:LLB}   
\end{figure*} 

\begin{figure*} 
\centerline{\vbox{
\centerline{\hbox{ 
\includegraphics[width=0.75\textwidth]{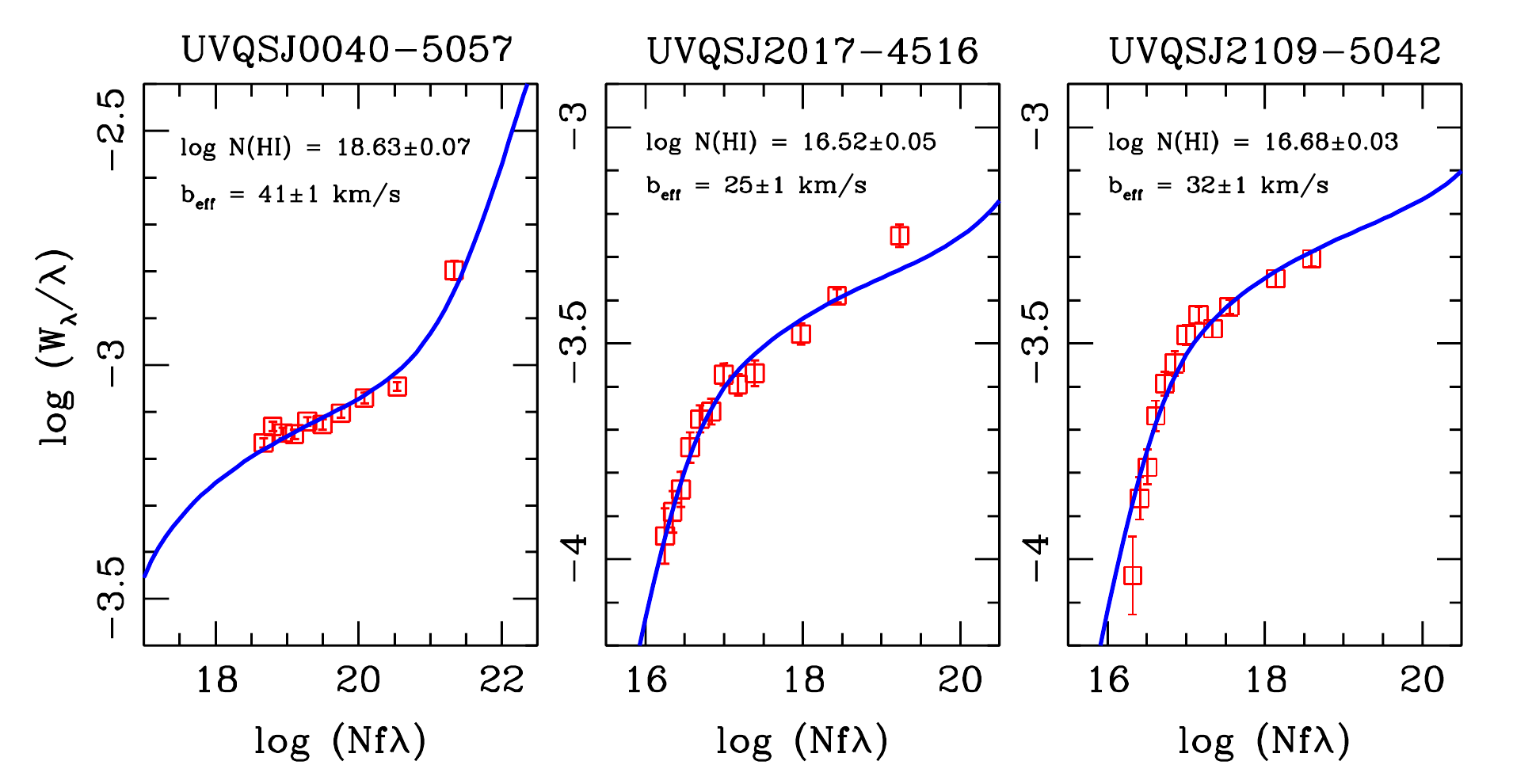}  
}} 
}}  
\vskip-0.2cm  
\caption{Results of the COG analysis of the Lyman series lines that are free from any significant contamination. The names of the quasars are given in the top. Note that the errors in the equivalent width measurements and hence in the inferred column densities are underestimated since the continuum placement uncertainties are not taken into account. An error of 0.10--0.15 dex in $\log N(\HI)$ is more reasonable. The \lya\ line from the \zabs~$=0.43737$ system towards UVQSJ0040--5057 falls on the damping part of the COG. In fact, we do see a weak damping wing in the \lya\ absorption.}        
\label{fig:cog}   
\end{figure*} 

\section{Analysis \& Results}    
\label{sec:analysis} 

In this section, we will first describe the properties of the absorbers and the targeted clusters. Next, the newly obtained data will be compared with those in the literature. \\

1. \underline{The \zabs~$=0.43737$ system towards UVQSJ0040--5057:} The absorber has a systemic velocity of $\approx-2600$~\kms\ with respect to the photometric-redshift (0.45$\pm$0.04) of the cluster J0041--5107 \citep[]{Bleem15}. This velocity is well within the $1\sigma$ uncertainty of the cluster redshift, i.e., $|\Delta z/(1+z)|\approx0.03\approx9000$~\kms. This is the strongest \HI\ absorber among the three systems studied here, producing a full \HI\ Lyman limit break at $\approx1315$~\AA\ (see the top panel of Fig.~\ref{fig:LLB}). The full break allows us to estimate a lower limit on $N(\HI)$\footnote{$N(\HI) = \frac{\tau_{\rm LL}}{\sigma_{\rm HI}}$; where $\sigma_{\rm HI}\approx$~6.3$\times10^{-18}$~cm$^{2}$ is the \HI\ photoionization cross-section and $\tau_{\rm LL}=-\ln (\frac{I}{I_{0}})$ is the optical depth at the Lyman limit.}  of $10^{17.7}$ \sqcm\ assuming the flux below $1315$~\AA\ to be less than $10^{-16}$ erg~cm$^{-2}$~s$^{-1}$~\AA$^{-1}$. A single component curve-of-growth (COG) analysis of all the unblended Lyman series lines yields a column density of $10^{18.63\pm0.07}$~\sqcm\ and an effective Doppler parameter of $b_{\rm eff}=$~41$\pm$1~\kms\ (see Fig.~\ref{fig:cog}). The mass of the corresponding cluster is $M_{500}=$~(3.04$\pm$0.87)$\times10^{14}$~\Msun, corresponding to $r_{500}=$~0.87~Mpc (see Table~\ref{tab:tab1}). The impact parameter of 3.80~Mpc gives $\rho_{\rm cl}/r_{500}\approx$~4.4. \smallskip

2. \underline{The \zabs$=0.43968$ system towards UVQSJ2017--4516:} The systemic velocity of the absorber with respect to the photometric redshift of the corresponding cluster (J2016--4517) is $\approx-2100$~\kms, which is well within the $1\sigma$ uncertainty ($|\Delta z/(1+z)|\approx0.02\approx6000$~\kms) of the cluster's photometric redshift \citep[]{Bleem15}. The absorber exhibits a partial break at the \HI\ Lyman limit corresponding to  $N(\HI)\approx10^{16.6}$~\sqcm\ (see middle panel of Fig.~\ref{fig:LLB}). A single component COG analysis of all the unblended Lyman series lines gives a consistent $N(\HI)$ of $10^{16.52\pm0.05}$~\sqcm\ and $b_{\rm eff}$ of 25$\pm$1~\kms\ (see Fig.~\ref{fig:cog}). The mass and radius of the cluster are $M_{500}=$~(3.19$\pm$0.89)$\times10^{14}$~\Msun\ and $r_{500}=0.89$~Mpc, respectively. The impact parameter of 4.20~Mpc corresponds to $\rho_{\rm cl}/r_{500}\approx$~4.7. \smallskip                

3. \underline{The \zabs$=0.51484$ system towards UVQSJ2109--5042:} This absorber shows a somewhat larger systemic velocity of $\approx+9000$~\kms\ with respect to the cluster J2109--5040. Such a velocity, however, is consistent within the $1\sigma$ uncertainty of the cluster's photometric redshift, i.e., $|\Delta z/(1+z)|$ $\approx0.03\approx9000$~\kms\ \citep[]{Bleem15}. The partial break seen in the COS spectrum (see the bottom panel of Fig.~\ref{fig:LLB}) gives $N(\HI) \approx10^{16.8}$~\sqcm\ which is consistent with the value we obtain from COG analysis of the unblended Lyman series lines (i.e., $10^{16.68\pm0.03}$~\sqcm, see Fig.~\ref{fig:cog}). We obtain $b_{\rm eff}$ of $32\pm1$~\kms\ from the COG analysis. The mass of the cluster, $M_{500}=$~(3.81$\pm$0.87)$\times10^{14}$~\Msun, corresponds to $r_{500}=0.93$~Mpc. The impact parameter of the cluster of 1.5~Mpc so that $\rho_{\rm cl}/r_{500}\approx$~1.6. \smallskip                   
%

\begin{figure} 
\centerline{\vbox{
\centerline{\hbox{ 
\includegraphics[width=0.50\textwidth]{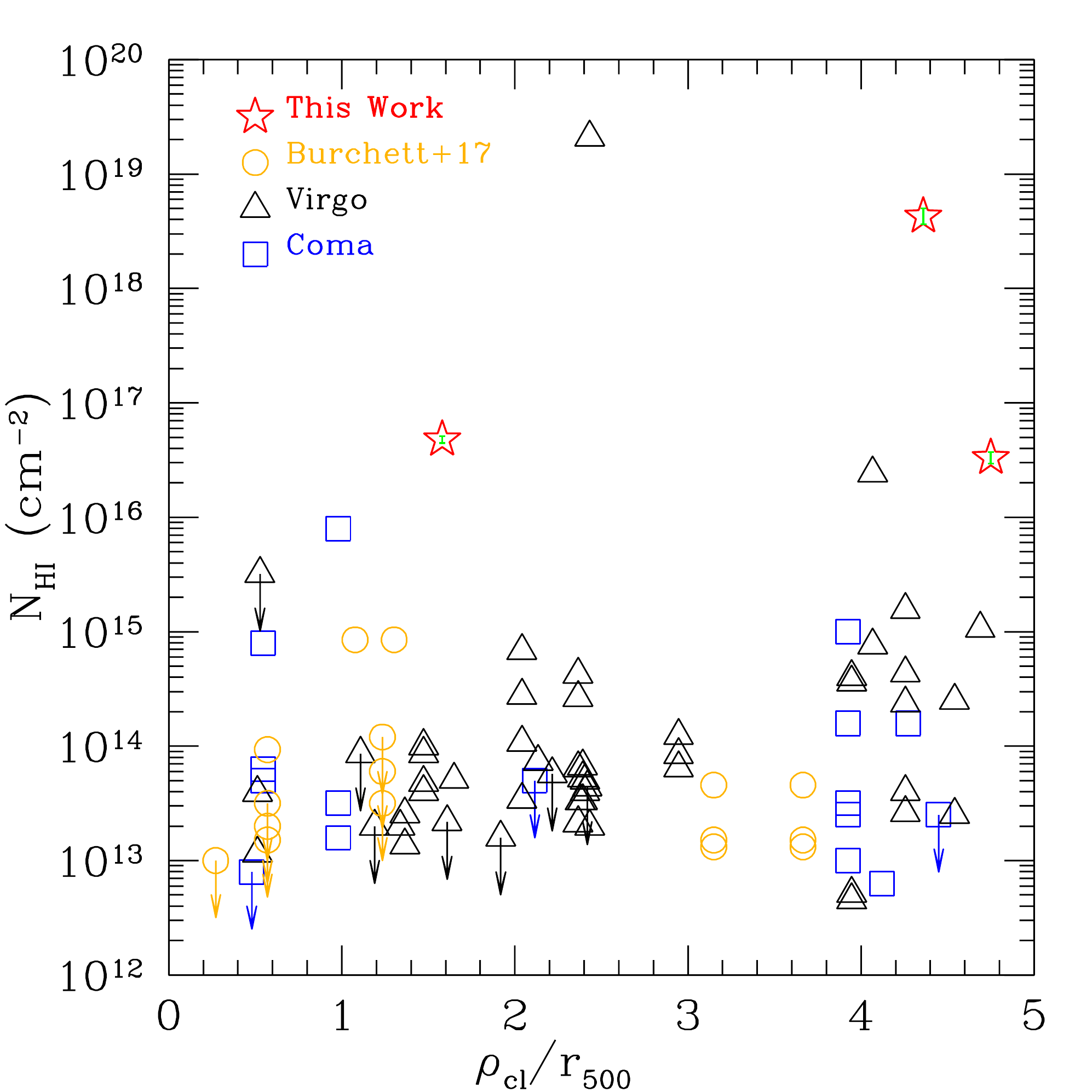}  
}} 
}}  
\caption{\HI\ column density against clustocentric impact parameter normalized by $r_{500}$. The data points corresponding to the Virgo and Coma clusters are from \citet{Yoon12} and \citet{Yoon17}. Since no convincing \lya\ absorption is seen within $\pm2000$~\kms\ of the cluster MaxBCG~J217.847+24.683\ in \citet{Burchett17}, we have assumed $N(\HI)<10^{13}$~\sqcm\ which is consistent with the error spectrum. No obvious trend is seen in the $N(\HI)$ radial profile.}         
\label{fig:clustocentric}     
\end{figure} 

In Fig.~\ref{fig:clustocentric}	we show the clustocentric radial profile of $N(\HI)$, combining our measurements with the handful of studies that exist in the literature. The data points corresponding to the Virgo and Coma clusters are taken from \citet{Yoon12} and \citet{Yoon17}. We have used the publicly available code $massconvert$ \citep[]{Hu03}, which assumes a NFW density profile, to convert the $M_{200}$ values of Coma ($1.4\times10^{15} M_{\odot}$) and Virgo ($2.2\times10^{14} M_{\odot}$) as given in \cite{Yoon17} to $M_{500}$ ($r_{500}$). The impact parameters of the quasar sightlines with respect to Virgo, which are not explicitly given in \cite{Yoon12}, are calculated assuming the center of the cluster to be the center of Cluster-A containing M87. Note that a velocity window of $\pm3024$~\kms\ around the systemic velocity of Coma and a velocity window of $-438$ to $+1862$~\kms\ around the systemic velocity of Virgo were considered by the authors for connecting absorbers to the corresponding cluster. The lower bound in velocity for Virgo was affected by the Galactic \lya\ absorption. In Fig.~\ref{fig:clustocentric} we also compare to the recent study of \cite{Burchett17} who have presented \HI\ column densities around 7 X-ray detected clusters in the redshift range 0.1--0.45 and within a velocity window of $\pm2000$~\kms. The $massconvert$ routine is used to convert their $M_{200}$ values to $M_{500}$ ($r_{500}$).     

The lack of any trend between $N(\HI)$ and the normalized clustocentric impact parameter is evident from Fig.~\ref{fig:clustocentric}. This is in contrast to the results of absorption line studies of the CGM \citep[e.g.,][]{Prochaska11,Tumlinson13} in which an anti-correlation between $N(\HI)$ (or equivalent width) and impact parameter is routinely seen. Next, we note that the SZ-effect selected clusters from this study are the ones that exhibit the highest \HI\ column densities in the outskirts. Only 3.7\% (2/54) of the Virgo-sightlines show a partial Lyman limit system (pLLS; $N(\HI)>10^{16.2}$~\sqcm). None of the Coma-sightlines show a pLLS. Note that, owing to the low redshifts, the $N(\HI)$ measurements for Virgo and Coma rely only on the \lya\ line and the systems with $N(\HI)>10^{14.5}$~\sqcm\ are presumably saturated. Nonetheless, even if we assume that all of the absorbers with $N(\HI)>10^{14.5}$~\sqcm\ towards Virgo and Coma are pLLS, the fraction is only \linebreak $\approx20\pm5$\% ($14/71$). None of the absorbers in \cite{Burchett17} sample are pLLS. The highest $N(\HI)$ they observed is $10^{14.93\pm0.03}$~\sqcm, which is well constrained by the presence of \lya, \lyb, and \lyg\ lines. All other absorbers have $\log N(\HI)/\rm cm^{-2} < 14.1$. In Section~\ref{sec:discuss}, we return to the issue of apparent abundance of strong \HI\ absorbers towards our targeted SZ-clusters.

\begin{figure} 
\centerline{\vbox{
\centerline{\hbox{ 
\includegraphics[width=0.50\textwidth]{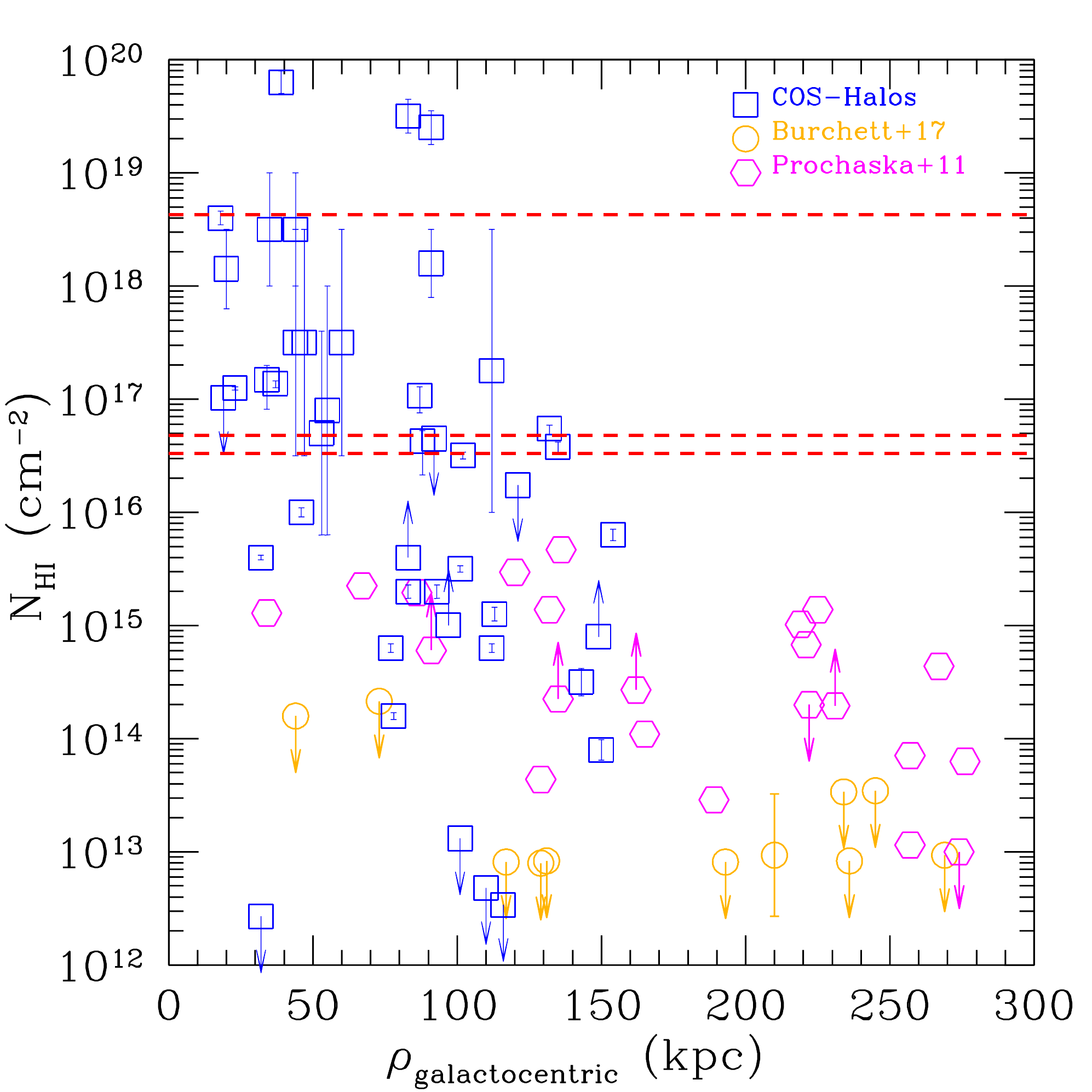}   
}} 
}}  
\caption{The radial profile of \HI\ absorbing gas around galaxies. The (blue) squares represent $L\sim L_*$ field galaxies at $z\approx0.2$ studied by the COS-Halos team \citep[]{Werk14,Prochaska17}. The (magenta) hexagons are from the sample of galaxies with $L>0.1L_*$ from \cite{Prochaska11}. Note that for each of these galaxies at least one additional galaxy with $L>0.1L_*$ has been detected within 3~Mpc of the sight line and with $|\Delta v| <400$~\kms. The (orange) circles represent the cluster galaxies from \cite{Burchett17}. The horizontal dashed lines are the $N(\HI)$ values for the three clusters studied here.}     
\label{fig:galactocentric}     
\end{figure}

In Fig.~\ref{fig:galactocentric} we show the \HI\ column density profile around low-$z$ galaxies using 3 different galaxy samples from the literature. The COS-Halos points represent the $N(\HI)$ profile around isolated, bright ($\sim L_*$) galaxies within $\pm600$~\kms. For the sample of \cite{Prochaska11} we have selected the galaxies with at least one additional $L>0.1L_*$ galaxy detected within 3 Mpc of the sightline and with velocity offset $|\Delta v|<400$~\kms. Therefore these data points essentially provide the radial profile of $N(\HI)$ around group galaxies. Here we note that some of the COS-Halos galaxies might also have companions within 3~Mpc and $|\Delta v|<400$~\kms. The points corresponding to \cite{Burchett17} are for cluster galaxies with $\rho_{\rm galactocentric}<300$~kpc and $|\Delta v|<600$~\kms. We note that at any given impact parameter, when data are available, the field galaxies tend to have the highest column densities followed by the group galaxies. As noted by \cite{Burchett17}, a significant suppression in $N(\HI)$ is apparent for their cluster galaxies. The $N(\HI)$ measurements of the SZ-clusters, as indicated by the dotted lines, are more than two orders of magnitude higher than the measurements/limits obtained by \cite{Burchett17}.

\section{Discussion \& Summary}    
\label{sec:discuss}

The \HI\ column densities we measure (i.e., $N(\HI)>10^{16.5}$~\sqcm) in the outskirts ($\rho_{\rm cl}/r_{500} = 1.6-4.7$) of SZ-selected clusters are clearly significantly higher than the other existing measurements around X-ray detected clusters (see Fig.~\ref{fig:clustocentric}). One of the possible reasons could be that the strong \HI\ absorbers we detect are not related to the targeted clusters. Recall that we noticed large velocity offsets (i.e., $-2600$, $-2100$, and $+9000$~\kms), albeit consistent within the $1\sigma$ uncertainties, between the absorber's and the cluster's redshifts. We, therefore, calculate the probability of random occurrence of each absorber around the cluster redshift within the $\pm1\sigma$ uncertainties in the photometric redshifts (see Table~\ref{tab:tab1}) by using the observed $\frac{d\mathcal{N}(>N, z)}{dz}$ of low-$z$ \HI\ absorbers \citep{Danforth16}. The probabilities turn out be $<4.2$\%, 6.3\%, and 8.7\% for the absorbers towards J0040--5057, J2017--4516, and J2109--5042 respectively. Since the events are independent, the compound probability of random occurrence of all three absorbers is $<0.02$\%. Thus it is {\em unlikely} that the absorbers are unrelated to the clusters. Nonetheless, spectroscopic confirmation of the redshifts of the clusters is of utmost importance.   

One of the interesting questions regarding the nature of the absorbers is: do they arise from the CCM or are they related to the CGM of cluster galaxies close to the lines of sight? If they are related to the CGM of $\sim L_*$ galaxies as in the COS-Halos survey, then it is evident from Fig.~\ref{fig:galactocentric} that the host-galaxies would be detected within 150~kpc of the QSO lines of sight. The majority of the absorbers with metal lines towards Virgo and Coma also have galaxies within 300~kpc and 300~\kms\ \citep[]{Yoon13,Yoon17}. Note that, we also detect a range of metal lines (e.g., \CII, \CIII, \NIII) from all of these absorbers. A faint dwarf galaxy very close to the QSO (possibly within the QSO PSF) can also produce such strong \HI\ absorption. Alternatively, these absorbers could be probing stripped-off CGM material far away from the host-galaxies. A uniform search for bright continuum-emitting and faint line-emitting galaxies in these fields, using facilities like the VLT/MUSE, is crucial for a better understanding of the origin(s) of the absorbers.  

Our single component COG analyses of the Lyman series lines yielded $b_{\rm eff}$ in the range of 25--41~\kms\ (Fig.~\ref{fig:cog}). This corresponds to a non-thermal broadening of 21--38~\kms\ assuming the temperature of the \HI\ absorbing  gas is $\sim10^{4}$~K. The presence of multiple (unresolved) components and/or higher gas temperature would only lessen the non-thermal contribution to the line broadening. In hydrodynamical simulations, the predicted merger-induced random gas motions are on the order a few 100~\kms\ in the outskirts of galaxy clusters \citep[see e.g., Fig. 4 in][]{Nagai13}. The measured $b_{\rm eff}$ values are consistent with the ones observed in the CGM of individual low-$z$ galaxies \citep[see e.g., Fig. 11 in][]{Tumlinson13} but are significantly lower than the values predicted in simulations of cluster outskirts. This suggests that the observed \HI\ absorbing gas is associated with the collapsed substructures as opposed to the CCM at large.     

If the absorbers stem from the CGM of cluster galaxies, then the difference of more than two orders of magnitude in $N(\HI)$ compared to the measurements/limits placed by \cite{Burchett17} is intriguing (see Fig.~\ref{fig:galactocentric}). We note that the clusters in \cite{Burchett17} are X-ray detected and primarily at $z\lesssim0.2$ but with masses ($\sim \rm few \times10^{14}$~\Msun), by and large, similar to our SZ-selected clusters at $z\approx0.46$. \linebreak It is unlikely that the redshift difference could make such a drastic difference in the observed $N(\HI)$ values. We further note that no unambiguous \HI\ absorption is detected (e.g., $N(\HI)<10^{14.3}$~\sqcm) from the highest redshift cluster galaxy ($z_{\rm gal}=$0.4560) towards the cluster GMBCG~J255.55+64.23\ in the sample of \cite{Burchett17}. This supports the idea that redshift evolution of cluster outskirts is unlikely to be the cause. However, currently we are restricted by a very small sample. UV spectroscopic observations of many more galaxy clusters are indispensable for further advancement.   

\cite{Yoon12} have noticed that the sightlines passing through substructures in the periphery of Virgo are likely to have higher \lya\ equivalent widths. The same scenario might be true for the systems studied here. In fact, a detection of a metal-poor ($\rm [O/H]=-1.6$), sub-damped \lya\ absorber (sub-DLA; $\log N(\HI)/\rm cm^{-2}\approx$~19.3) has been reported by \cite{Tripp05a} in the outskirts of Virgo near the NGC 4261 galaxy group (see Fig.~\ref{fig:clustocentric}). Interestingly, no bright galaxy with a small impact parameter has been found by the authors. The nearest known sub-$L_*$ and $L_*$ galaxies have impact parameters of $\approx$~90~kpc and $\approx$~260~kpc respectively. From the observed low metallicity, the underabundance of nitrogen, and the lack of $\alpha-$element enhancement the authors argued that the absorber is related to a dwarf galaxy and/or a high velocity cloud in the outskirts of Virgo. The strongest absorber in the sample of \cite{Burchett17}, with $\log N(\HI)/\rm cm^{-2}\approx$~14.9, also does not have any bright galaxy counterpart within 300~kpc and $|\Delta v|<$ 400~\kms. In this case, the sightline passes through the interface of sub-clusters A1095W and A1095E. The authors have suggested that stripping from a far away galaxy or a density enhancement due to a merger shockwave are possible origins for the cool gas detected in absorption.      

In addition to strong \HI, all three systems exhibit strong absorption lines from low- (e.g., \CII$\lambda1036$) and intermediate- (e.g., \CIII$\lambda$977, \NIII$\lambda$989) ionization metal lines, suggesting high metallicity gas. Weak high-ionization lines (e.g., \OVI, \NeVIII) might also be present in at least one of them. A detailed analysis of the metal lines, along with ionization models, will be presented in future papers. The analysis of the detected metal lines will provide further insights into the nature of the absorbers.

\acknowledgments
Support for this research was provided by NASA through grants HST GO-14655 from the Space Telescope Science Institute, which is operated by the Association of Universities for Research in Astronomy, Inc., under NASA contract NAS5-26555. SM and JS acknowledge support from European Research Council (ERC), Grant Agreement 278594-GasAroundGalaxies.


\begin{thebibliography}{}
\expandafter\ifx\csname natexlab\endcsname\relax\def\natexlab#1{#1}\fi
\providecommand{\url}[1]{\href{#1}{#1}}

\bibitem[{{Bah{\'e}} {et~al.}(2017){Bah{\'e}}, {Barnes}, {Dalla Vecchia},
  {Kay}, {White}, {McCarthy}, {Schaye}, {Bower}, {Crain}, {Theuns}, {Jenkins},
  {McGee}, {Schaller}, {Thomas}, \& {Trayford}}]{Bahe17}
{Bah{\'e}}, Y.~M., {Barnes}, D.~J., {Dalla Vecchia}, C., {et~al.} 2017, ArXiv
  e-prints, arXiv:1703.10610

\bibitem[{{Bleem} {et~al.}(2015){Bleem}, {Stalder}, {de Haan}, {Aird}, {Allen},
  {Applegate}, {Ashby}, {Bautz}, {Bayliss}, {Benson}, {Bocquet}, {Brodwin},
  {Carlstrom}, {Chang}, {Chiu}, {Cho}, {Clocchiatti}, {Crawford}, {Crites},
  {Desai}, {Dietrich}, {Dobbs}, {Foley}, {Forman}, {George}, {Gladders},
  {Gonzalez}, {Halverson}, {Hennig}, {Hoekstra}, {Holder}, {Holzapfel},
  {Hrubes}, {Jones}, {Keisler}, {Knox}, {Lee}, {Leitch}, {Liu}, {Lueker},
  {Luong-Van}, {Mantz}, {Marrone}, {McDonald}, {McMahon}, {Meyer}, {Mocanu},
  {Mohr}, {Murray}, {Padin}, {Pryke}, {Reichardt}, {Rest}, {Ruel}, {Ruhl},
  {Saliwanchik}, {Saro}, {Sayre}, {Schaffer}, {Schrabback}, {Shirokoff},
  {Song}, {Spieler}, {Stanford}, {Staniszewski}, {Stark}, {Story}, {Stubbs},
  {Vanderlinde}, {Vieira}, {Vikhlinin}, {Williamson}, {Zahn}, \&
  {Zenteno}}]{Bleem15}
{Bleem}, L.~E., {Stalder}, B., {de Haan}, T., {et~al.} 2015, \apjs, 216, 27

\bibitem[{{Burchett} {et~al.}(2017){Burchett}, {Tripp}, {Wang}, {Willmer},
  {Bowen}, \& {Jenkins}}]{Burchett17}
{Burchett}, J.~N., {Tripp}, T.~M., {Wang}, Q.~D., {et~al.} 2017, ArXiv
  e-prints, arXiv:1705.05892

\bibitem[{{Danforth} {et~al.}(2010){Danforth}, {Stocke}, \&
  {Shull}}]{Danforth10}
{Danforth}, C.~W., {Stocke}, J.~T., \& {Shull}, J.~M. 2010, \apj, 710, 613

\bibitem[{{Danforth} {et~al.}(2016){Danforth}, {Keeney}, {Tilton}, {Shull},
  {Stocke}, {Stevans}, {Pieri}, {Savage}, {France}, {Syphers}, {Smith},
  {Green}, {Froning}, {Penton}, \& {Osterman}}]{Danforth16}
{Danforth}, C.~W., {Keeney}, B.~A., {Tilton}, E.~M., {et~al.} 2016, \apj, 817,
  111

\bibitem[{{Gonzalez} {et~al.}(2013){Gonzalez}, {Sivanandam}, {Zabludoff}, \&
  {Zaritsky}}]{Gonzalez13}
{Gonzalez}, A.~H., {Sivanandam}, S., {Zabludoff}, A.~I., \& {Zaritsky}, D.
  2013, \apj, 778, 14

\bibitem[{{Gonzalez} {et~al.}(2007){Gonzalez}, {Zaritsky}, \&
  {Zabludoff}}]{Gonzalez07}
{Gonzalez}, A.~H., {Zaritsky}, D., \& {Zabludoff}, A.~I. 2007, \apj, 666, 147

\bibitem[{{Green} {et~al.}(2012){Green}, {Froning}, {Osterman}, {Ebbets},
  {Heap}, {Leitherer}, {Linsky}, {Savage}, {Sembach}, {Shull}, {Siegmund},
  {Snow}, {Spencer}, {Stern}, {Stocke}, {Welsh}, {B{\'e}land}, {Burgh},
  {Danforth}, {France}, {Keeney}, {McPhate}, {Penton}, {Andrews},
  {Brownsberger}, {Morse}, \& {Wilkinson}}]{Green12}
{Green}, J.~C., {Froning}, C.~S., {Osterman}, S., {et~al.} 2012, \apj, 744, 60

\bibitem[{{Hu} \& {Kravtsov}(2003)}]{Hu03}
{Hu}, W., \& {Kravtsov}, A.~V. 2003, \apj, 584, 702

\bibitem[{{Kacprzak} {et~al.}(2015){Kacprzak}, {Muzahid}, {Churchill},
  {Nielsen}, \& {Charlton}}]{Kacprzak15}
{Kacprzak}, G.~G., {Muzahid}, S., {Churchill}, C.~W., {Nielsen}, N.~M., \&
  {Charlton}, J.~C. 2015, \apj, 815, 22

\bibitem[{{Lau} {et~al.}(2015){Lau}, {Nagai}, {Avestruz}, {Nelson}, \&
  {Vikhlinin}}]{Lau15}
{Lau}, E.~T., {Nagai}, D., {Avestruz}, C., {Nelson}, K., \& {Vikhlinin}, A.
  2015, \apj, 806, 68

\bibitem[{{Monroe} {et~al.}(2016){Monroe}, {Prochaska}, {Tejos}, {Worseck},
  {Hennawi}, {Schmidt}, {Tumlinson}, \& {Shen}}]{Monroe16}
{Monroe}, T.~R., {Prochaska}, J.~X., {Tejos}, N., {et~al.} 2016, \aj, 152, 25

\bibitem[{{Nagai} {et~al.}(2013){Nagai}, {Lau}, {Avestruz}, {Nelson}, \&
  {Rudd}}]{Nagai13}
{Nagai}, D., {Lau}, E.~T., {Avestruz}, C., {Nelson}, K., \& {Rudd}, D.~H. 2013,
  \apj, 777, 137

\bibitem[{{Osterman} {et~al.}(2011){Osterman}, {Green}, {Froning},
  {B{\'e}land}, {Burgh}, {France}, {Penton}, {Delker}, {Ebbets}, {Sahnow},
  {Bacinski}, {Kimble}, {Andrews}, {Wilkinson}, {McPhate}, {Siegmund}, {Ake},
  {Aloisi}, {Biagetti}, {Diaz}, {Dixon}, {Friedman}, {Ghavamian}, {Goudfrooij},
  {Hartig}, {Keyes}, {Lennon}, {Massa}, {Niemi}, {Oliveira}, {Osten},
  {Proffitt}, {Smith}, \& {Soderblom}}]{Osterman11}
{Osterman}, S., {Green}, J., {Froning}, C., {et~al.} 2011, \apss, 335, 257

\bibitem[{{Prochaska} {et~al.}(2011){Prochaska}, {Weiner}, {Chen}, {Mulchaey},
  \& {Cooksey}}]{Prochaska11}
{Prochaska}, J.~X., {Weiner}, B., {Chen}, H.-W., {Mulchaey}, J., \& {Cooksey},
  K. 2011, \apj, 740, 91

\bibitem[{{Prochaska} {et~al.}(2017){Prochaska}, {Werk}, {Worseck}, {Tripp},
  {Tumlinson}, {Burchett}, {Fox}, {Fumagalli}, {Lehner}, {Peeples}, \&
  {Tejos}}]{Prochaska17}
{Prochaska}, J.~X., {Werk}, J.~K., {Worseck}, G., {et~al.} 2017, \apj, 837, 169

\bibitem[{{Simionescu} {et~al.}(2011){Simionescu}, {Allen}, {Mantz}, {Werner},
  {Takei}, {Morris}, {Fabian}, {Sanders}, {Nulsen}, {George}, \&
  {Taylor}}]{Simionescu11}
{Simionescu}, A., {Allen}, S.~W., {Mantz}, A., {et~al.} 2011, Science, 331,
  1576

\bibitem[{{Tripp} {et~al.}(2005){Tripp}, {Jenkins}, {Bowen}, {Prochaska},
  {Aracil}, \& {Ganguly}}]{Tripp05a}
{Tripp}, T.~M., {Jenkins}, E.~B., {Bowen}, D.~V., {et~al.} 2005, \apj, 619, 714

\bibitem[{{Tumlinson} {et~al.}(2013){Tumlinson}, {Thom}, {Werk}, {Prochaska},
  {Tripp}, {Katz}, {Dav{\'e}}, {Oppenheimer}, {Meiring}, {Ford}, {O'Meara},
  {Peeples}, {Sembach}, \& {Weinberg}}]{Tumlinson13}
{Tumlinson}, J., {Thom}, C., {Werk}, J.~K., {et~al.} 2013, \apj, 777, 59

\bibitem[{{Turner} {et~al.}(2014){Turner}, {Schaye}, {Steidel}, {Rudie}, \&
  {Strom}}]{Turner14}
{Turner}, M.~L., {Schaye}, J., {Steidel}, C.~C., {Rudie}, G.~C., \& {Strom},
  A.~L. 2014, \mnras, 445, 794

\bibitem[{{Urban} {et~al.}(2014){Urban}, {Simionescu}, {Werner}, {Allen},
  {Ehlert}, {Zhuravleva}, {Morris}, {Fabian}, {Mantz}, {Nulsen}, {Sanders}, \&
  {Takei}}]{Urban14}
{Urban}, O., {Simionescu}, A., {Werner}, N., {et~al.} 2014, \mnras, 437, 3939

\bibitem[{{Voit}(2005)}]{Voit05}
{Voit}, G.~M. 2005, Reviews of Modern Physics, 77, 207

\bibitem[{{Walker} {et~al.}(2012){Walker}, {Fabian}, {Sanders}, \&
  {George}}]{Walker12}
{Walker}, S.~A., {Fabian}, A.~C., {Sanders}, J.~S., \& {George}, M.~R. 2012,
  \mnras, 427, L45

\bibitem[{{Werk} {et~al.}(2014){Werk}, {Prochaska}, {Tumlinson}, {Peeples},
  {Tripp}, {Fox}, {Lehner}, {Thom}, {O'Meara}, {Ford}, {Bordoloi}, {Katz},
  {Tejos}, {Oppenheimer}, {Dav{\'e}}, \& {Weinberg}}]{Werk14}
{Werk}, J.~K., {Prochaska}, J.~X., {Tumlinson}, J., {et~al.} 2014, \apj, 792, 8

\bibitem[{{Yoon} \& {Putman}(2013)}]{Yoon13}
{Yoon}, J.~H., \& {Putman}, M.~E. 2013, \apjl, 772, L29

\bibitem[{{Yoon} \& {Putman}(2017)}]{Yoon17}
---. 2017, \apj, 839, 117

\bibitem[{{Yoon} {et~al.}(2012){Yoon}, {Putman}, {Thom}, {Chen}, \&
  {Bryan}}]{Yoon12}
{Yoon}, J.~H., {Putman}, M.~E., {Thom}, C., {Chen}, H.-W., \& {Bryan}, G.~L.
  2012, \apj, 754, 84

\end{thebibliography}
\end{document}